\documentclass[sigconf]{aamas}

\usepackage{balance} 
\usepackage[T1]{fontenc}
\usepackage{graphicx} 
\usepackage{float}
\usepackage{booktabs} 
\usepackage[ruled]{algorithm2e} 
\usepackage{diagbox}
\usepackage{booktabs}
\usepackage{amsmath}
\usepackage{placeins}

\usepackage{tcolorbox}

\definecolor{gamecolor}{rgb}{0.8, 0.8, 0.8}
\newtcolorbox{game}[1]{colback=gamecolor!5!white,colframe=gamecolor!75!black,fonttitle=\bfseries,title=#1}

\newcommand{\mathbbm}[1]{\text{\usefont{U}{bbm}{m}{n}#1}} 

\newcommand{\numplayers}{N}

\setcopyright{ifaamas}
\acmConference[AAMAS '25]{Workshop on Autonomous Robots and Multirobot Systems (ARMS), at Autonomous Agents and Multiagent Systems (AAMAS 2025)}{May 19 -- 23, 2025}
{Detroit, Michigan, USA}{}
\copyrightyear{2025}
\acmYear{2025}
\acmDOI{}
\acmPrice{}
\acmISBN{}

\acmSubmissionID{<<submission id>>}

\title[Mixed Strategy Constraints in Continuous Games]{Mixed Strategy Constraints in Continuous Games}

\author{Mel Krusniak}
\affiliation{
  \institution{Vanderbilt University}
  \city{Nashville, TN}
  \country{USA}}
\email{mel.krusniak@vanderbilt.edu}

\author{Forrest Laine}
\affiliation{
  \institution{Vanderbilt University}
  \city{Nashville, TN}
  \country{USA}}
\email{forrest.laine@vanderbilt.edu}

\begin{abstract}
When modeling robot interactions as Nash equilibrium problems, it is desirable to place coupled constraints which restrict these interactions to be safe and acceptable (for instance, to avoid collisions). Such games are continuous with potential mixed strategy equilibria, and this combination of characteristics means special care must be given to setting coupled constraints in a way that respects mixed strategies while remaining compatible with continuous game solution methods. Here, we investigate the problem of constraint-setting in this context, primarily focusing on a chance-based method. We first motivate these chance constraints in a discrete setting, placing them on $n$-player matrix games as a justifiable approach to handling the probabilistic nature of mixing. Then, we describe a numerical solution method for these chance constrained, continuous games with simultaneous pure strategy optimization. Finally, using a modified pursuit-evasion game as a motivating example, we demonstrate the actual behavior of this solution method in terms of its fidelity, parameter sensitivity, and efficiency.
\end{abstract}

\keywords{Game theory, mixed strategies, constrained optimization}

\newcommand{\BibTeX}{\rm B\kern-.05em{\sc i\kern-.025em b}\kern-.08em\TeX}

\begin{document}


\pagestyle{fancy}
\fancyhead{}


\maketitle 


\section{Introduction}

When designing models of multi-agent interactions in robotics and control, it is frequently necessary to place coupled constraints on agent behavior --- for instance, preventing agents from colliding or jointly disrupting the environment. Competitive control scenarios are often described as continuous normal form games, but when coupled constraints are present, a choice is required about how to implement the constraints when many potential strategy profiles might be realized. In this work, we discuss proper engineering of constraints in a mixed-strategy, continuous framework, and put forth a chance-constraint-based approach as a compelling method.

We begin by discussing the behavior of coupled constraints in discrete games with mixed strategies. Since mixed strategies are probabilistic, a constraint may or may not be violated depending on the strategies actually realized by all players, and we describe the main intuitive methods used to account for this. Next, we consider continuous games. In general, continuous games require mixing over an infinite support of pure strategies to ensure the existence of a Nash equilibrium, but in certain classes of games such as separable games, finite support mixed strategy equilibria exist (\cite{stein2008separable}), and constraints can be placed across these finite strategies. However, solver approaches for continuous games require the constraints to be differentiable, introducing another consideration \cite{peters2022lifting}. To that end, we develop an iterative-tightening solver approach allowing coupled chance constraints to be used in continuous games.

To provide a concrete example, we consider a constrained, continuous, three-player pursuit-evasion game. We demonstrate the behavior of the chance-based constraint-setting approach (and an alternative), and analyze efficacy in maintaining the constraint, permissiveness of mixed strategy reasoning, and potential unintended consequences of the modeling choice. We conclude with promising areas of further investigation, and a discussion of the application of these constraints to model common problems in our spatial domain.

\subsection{Our Contributions}

This work presents the following contributions:

\begin{itemize}
    \item A formulation of chance constraints applicable to mixed-strategy ``tensor games with tensor constraints'' (normal form games with coupled constraints);
    \item A numerical solver which uses this formulation to find mixed-strategy Nash equilibria in continuous games with an iterative-tightening approach; and
    \item Empirical behavior of this modeling approach (and an alternative) in a constrained pursuit-evasion game, including analysis of quality with respect to its parameters and the feasibility of the resulting solution.
\end{itemize}

The structure of this paper is as follows: In the remainder of this section we briefly discuss the related work on generalized Nash equilibrium problems and chance constraints to scope our work. In Section 2 we discuss ``tensor games with tensor constraints'' --- a convenient set of normal form games with coupled constraints --- and potential constraint-setting approaches. In Section 4, we move to the continuous setting and describe a solver with which we compute the empirical results discussed in Section 5. We close with a brief consideration of practical considerations and future work.

\subsection{Related Work}
\subsubsection{Coupled constraints}
In computational game theory, \textit{coupled constraints} are constraints which involve the decision variables of multiple players. These are particularly important in multi-agent robotics and control, where they prevent collisions or other undesirable interactions. Games which feature them are known as \textit{generalized Nash equilibrium problems}, or GNEPs \cite{facchinei2010generalized}. Older terminology refers to coupled constraints as \textit{feasible set interactions}, found in \textit{social equilibrium problems} and \textit{pseudo-Nash games}  \cite{harker1991generalized}. 

Solutions to solving GNEPs vary, and do not typically use an explicit concept of mixed strategies. \cite{krawczyk2007numerical} summarizes numerical solutions to solving these games, a project further advanced in, e.g., \cite{von2009numerical}. Past avenues include penalty methods \cite{facchinei2010penalty}, augmented Lagrangian methods \cite{kanzow2016augmented}, and methods informed by error bound theory \cite{ba2022exact}. This work focuses on the representation and modeling of constraints in mixed-strategy games, and less on computation. In this work we characterize GNEPs as a mixed complementarity problem via the KKT conditions, and solve them via a readily available solver \cite{dirkse1995path}.

We diverge somewhat from this body of work in that we do explicitly consider mixed strategies, optimizing both pure strategies and mixing weights to solve continuous games. To our knowledge, constraint design in this form of generalized Nash equilibrium has not yet been fully explored, which prevents further development of mixed-strategy agents in robotics and control. 

\subsubsection{Chance constraints}
To intuitively account for the stochasticity of mixed strategies, we use chance constraints, originally formulated in \cite{charnes1959chancedef}. Chance-constrained problems without feasible set interactions (that is, in ordinary Nash equilibrium problems) are well covered, and significant work has been done to characterize the circumstances in which such problems have equilibria (as in \cite{henrion2008chanceconvexity}, \cite{geletu2013chanceapplications}). Chance constraints have been used in the presence of mixing for decades, see e.g. \cite{mukherjee1980mixed}.

Some work has been done describing the behavior of chance constraints in GNEPs. \cite{zhang2022variational} manage mixing chance constraints in discrete mixed-strategy scenarios from the variational inequality point of view, while \cite{singh2018variational} apply the same treatment to a continuous single strategy to handle stochastic payoffs, but the former assumes preselected pure strategies, while the latter does not mix. We assume pure strategy parameters are \textit{additional} continuous decision variables rather than as given, and mix over them under the presence of chance constraints.

\section{Formulation}
In this section, we discuss discrete Nash equilibrium problems with coupled constraints --- by definition, existence of equilibria, and potential constraint-setting approaches.

\subsection{Tensor games with tensor constraints}
\label{sec:boilerplate}
We begin by describing a set of ``generalized'' normal form games (generalized in the sense of ``generalized Nash equilibrium problems''; that is, normal form games with coupled constraints). This is not a novel formulation, but it is a convenient one for our purposes.

\newcommand{\reals}{\mathbb{R}}
\newcommand{\bools}{\{ 0, 1 \}}
\newcommand{\expc}{\mathbb{E}}

\newcommand{\vartensorgame}{G}
\newcommand{\varcosttensor}{A}
\newcommand{\varconstrainttensor}{Q}

\newcommand{\varconfidence}{\epsilon}

\newcommand{\varprobs}{x}

\newcommand{\varstratsize}{d}

\newcommand{\numstrats}{m}
\newcommand{\numconstraints}{C}

\newcommand{\funcost}{f}
\newcommand{\funconstraint}{g}

\newcommand{\idxtensor}{K}
\newcommand{\idxplayer}{i}
\newcommand{\idxconstraint}{j}
\newcommand{\idxstrat}{k}
\newcommand{\prmstrictness}{\omega}

\newcommand{\setownership}{\mathcal{O}}
\newcommand{\setdiscretestrats}{\mathcal{S}}

\newcommand{\varstrats}{s}
\newcommand{\funsoftener}{\rho}

\newcommand{\oprtprod}{\;[\cdot]\;}

We work with normal form games as expressed with tensors, sometimes referred to as \textit{tensor games} in computational game theory. Each player $\idxplayer$ mixes over $\numstrats_\idxplayer$ pure strategies with probabilities $\varprobs_\idxplayer$. We denote player $\idxplayer$'s $\idxstrat$-th pure strategy $\varstrats_{\idxplayer, \idxstrat}$ and that player's pure strategy space as $\setdiscretestrats_\idxplayer$, such that $\varstrats_\idxplayer \in \setdiscretestrats_\idxplayer^{\numstrats_\idxplayer}$ (and similarly denoting $\varstrats \in \setdiscretestrats$).

Each player has a \textit{cost tensor}, and each element of the cost tensor corresponds to a joint strategy. Cost tensors are denoted $\varcosttensor_\idxplayer$, and all $\varcosttensor_\idxplayer \in \reals^{m_1, .., m_\idxplayer}$. The resulting cost, when each player has probability $\varprobs_{\idxplayer, \idxstrat}$ assigned to their $\idxstrat$th strategy, is a homogeneous $\numplayers$-polynomial. Lacking more familiar notation, we denote the operation that forms this polynomial $\varcosttensor_\idxplayer \oprtprod \varprobs$, under the following definition:

\begin{definition} 
    For a tensor $T$ and vectors $x_1 .. x_N$, 
    $T \oprtprod x$ denotes the dot product of every vector $x_i$ along the $\idxplayer$th axis of $T$. That is,
    \begin{equation}
        T \oprtprod \varprobs := (((T \cdot \varprobs_\idxplayer) \cdot ..) \cdot \varprobs_\numplayers
        \label{eqn:tprod_def}
    \end{equation}
\end{definition}

We also constrain these games with a total of $\numconstraints$ constraints (indexed by $\idxconstraint)$. Each player may or may not respect a particular constraint; we denote the index set of constraints respected by player $\idxplayer$ as $\setownership_\idxplayer$. Each constraint tensor is denoted as $\varconstrainttensor_\idxconstraint$ and has the same shape as the cost tensor. The elements of the constraint tensor represent some value regarding the feasibility of the joint strategy with the corresponding index, and we constrain the expected value: $\varconstrainttensor_\idxconstraint \oprtprod \varprobs \geq \varconfidence_\idxconstraint$ for some arbitrary $\varconfidence_\idxconstraint$.  

With this notation in place, we define ``tensor games with tensor constraints'':

\begin{definition}
    A \textit{tensor game with tensor constraints} $\vartensorgame(\varcosttensor, \varconstrainttensor, \epsilon)$ takes the following form, given the cost tensors $\varcosttensor_\idxplayer$, constraint tensors $\varconstrainttensor_{i,\idxconstraint}$, and thresholds $\epsilon_{i,j}$:

    \begin{equation} \label{eq:tgtc}
        \forall \idxplayer \in \{1 .. \numplayers\}: \begin{bmatrix}
            \;\;\min\limits_{\varprobs_\idxplayer \in \Delta^{\numstrats_{\idxplayer - 1}}}\;\; \varcosttensor_\idxplayer \oprtprod \varprobs \;\;\;\; \textrm{s.t.}\;\;\;\;\\[8pt]
            \;\;\;\;\;\;\varconstrainttensor_{i,\idxconstraint} \oprtprod x - \varconfidence_{i,\idxconstraint} \geq 0 \;\; \forall \idxconstraint \in \setownership_i
        \end{bmatrix}
    \end{equation}
    \label{eqn:def_tensor_game}
\end{definition}

Let $x_{\neg i}$ denote the concatenation of strategy vectors 
$$x_1,...,x_{i-1},x_{i+1},...,x_N$$
and let $(x_i,x_{\neg i})$ be shorthand for $x$. To be entirely explicit, we define Nash equilibrium on tensor games with tensor constraints:

\begin{definition}
    A \textit{Nash equilibrium} for a tensor game with tensor constraints $\vartensorgame(\varcosttensor, \varconstrainttensor, \epsilon)$ is a strategy vector $x^*$ such that for all $i \in \{1,...,N\}$,
    \begin{equation}
    \begin{split}
        A_i \oprtprod (x_i^*,x_{\neg i}^*) \le A_i \oprtprod (x_i, x_{\neg i}^*)  \\ \forall x_i \in \Delta^{m_{i}-1} : \varconstrainttensor_{i,\idxconstraint} \oprtprod (x_i, x_{\neg i}^*) - \epsilon_{i,j} \ge 0 \; \; \forall j \in \setownership_i
    \end{split}
    \end{equation}
\end{definition}

In discrete games, the way in which the contents of the cost and constraint tensors are derived is external to the game. However, to model continuous games, we assume that the elements of the tensors are related to a set pure strategies $s$ by way of real-valued cost and constraint functions $\funcost_\idxplayer$ and $\funconstraint_{i,\idxconstraint}: \setdiscretestrats \rightarrow \reals$. (We implicitly assume players also choose a finite number of pure strategies from $\mathcal{S}$ which may or may not support a Nash equilibrium. We discuss this approach in greater detail Section \ref{sec:implementation}.)

Let $\idxtensor$ be an index into a cost or constraint tensor (and therefore $\idxtensor_{\idxplayer}$ is the index of the strategy realized by player $\idxplayer$ in a particular joint strategy). Then, the cost and constraint tensor elements are defined accordingly:

\begin{equation}
    \varcosttensor_{\idxplayer, \idxtensor} = \funcost_\idxplayer(\varstrats_{1, \idxtensor_1} \;..\; \varstrats_{\numplayers, \idxtensor_\numplayers}) \;\; ; \;\;
    \varconstrainttensor_{\idxplayer, j, \idxtensor} = \funsoftener(\funconstraint_{\idxplayer,j}(\varstrats_{1, \idxtensor_1} \;..\; \varstrats_{\numplayers, \idxtensor_\numplayers}))  
    \label{eqn:def_lift_tensor_elements}
\end{equation}

We use the convention that $\funconstraint_{i,\idxconstraint}(s_\idxtensor) \geq 0$ indicates a satisfied constraint (that is, a feasible joint strategy).

When solving a continuous tensor game with tensor constraints, our goal (from the perspective of player $\idxplayer$) is to both mix over pure strategies correctly, and to select $\varstrats_\idxplayer$ that generate the most favorable $\varcosttensor$ and $\varconstrainttensor$ under a correct mix. For this purpose we assume $f$ and $g$ are twice differentiable and defined over a compact $\setdiscretestrats$.

The function $\funsoftener_\idxconstraint$ is the \textit{constraint indicating function}. In theory, $\funsoftener(\funcost(\varstrats_\idxtensor)) = \mathbbm{1}(\funcost(\varstrats_\idxtensor) \geq 0)$, but we will consider examples where this is not the case.

\subsection{Existence of equilibria}
Tensor games with tensor constraints do not necessarily admit a Nash equilibrium. As such, we briefly pause to establish sufficient conditions to ensure the presence of an equilibrium in a tensor-constrained tensor game. 

\begin{theorem}
    Tensor games with tensor constraints have at least one equilibrium if, for all sets of opponent strategies $\varprobs_{\neg \idxplayer}$, there is a feasible ego strategy $\varprobs_\idxplayer$, and an open set of feasible strategies around it. 
    \label{thm:constrained_tgs_have_equilibria}
\end{theorem}

\newcommand{\setconstraintcorrespondance}{C}
\newcommand{\funcostouter}{\theta}

\begin{proof}
     Using the notation from Eq. \ref{eqn:def_tensor_game} and Eq. $\ref{eqn:def_lift_tensor_elements}$, let $\setconstraintcorrespondance_\idxplayer(\varprobs_{\neg \idxplayer}) := \{\varprobs_\idxplayer: \forall \idxconstraint \in \setownership_\idxplayer: \varconstrainttensor_{i,\idxconstraint} \oprtprod \varprobs - \varconfidence_{i,\idxconstraint} \geq 0\}$ denote the constraint correspondance from joint opponent strategies to the set of feasible ego strategies. Let $\funcostouter(\varprobs) := \varcosttensor^{(\idxplayer)} \oprtprod \varprobs$ denote the overall cost function for each player.

     We then follow the conditions put forth by \cite{dutang2013existence}:
    \begin{enumerate}
        \item $\Delta^{\numstrats_\idxplayer - 1}$, a simplex, is a non-empty, compact, convex polytope.
        
        \item $\varcosttensor_\idxplayer \oprtprod \varprobs$ is continuous on graph $\textrm{Gr}(\setconstraintcorrespondance_\idxplayer(\varprobs_{\neg \idxplayer}))$ because it is a multilinear map, which is continuous on the entire joint strategy space.

        \item $\varprobs_\idxplayer \rightarrow \funcostouter_\idxplayer(\varprobs_\idxplayer; \varprobs_{\neg \idxplayer})$ is quasiconcave on $\setconstraintcorrespondance_\idxplayer(\varprobs_{\neg \idxplayer})$ for all $\varprobs_\idxplayer\in \Delta^{\numstrats_\idxplayer - 1}$ because $\funcostouter_\idxplayer$ is a multilinear map and therefore linear given $\varprobs_{\neg \idxplayer}$.

        \item $\setconstraintcorrespondance_\idxplayer(\varprobs_{\neg \idxplayer})$ is nonempty, closed and convex for all $\varprobs_{\neg \idxplayer} \in \Delta^{\numstrats_{\neg \idxplayer} - 1}$. This follows from the assumption that each player has an open set of feasible strategies given any opponent strategies.

        \item First, $\varconstrainttensor_\idxconstraint \oprtprod \varprobs$ is continuous and concave in $\varprobs_\idxplayer$ given all $\varprobs_{\neg \idxplayer}$, since it is linear in that case. Second, we have assumed that $\setconstraintcorrespondance_\idxplayer(\varprobs_{\neg \idxplayer})$ contains some nonempty open set of feasible solutions. These two facts satisfy Prop. 4.2 of \cite{dutang2013existence}, which guarantees lower and upper semicontinuity of $\setconstraintcorrespondance_\idxplayer$ in $\varprobs_{\neg \idxplayer}$
    \end{enumerate}

    With these conditions satisfied, tensor games with tensor constraints satisfying the assumption must have at least one equilibrium.
\end{proof}

Note that the open-set assumption is sufficient, but not necessary; there are tensor games with equilibria that do not satisfy it.

\subsection{As mixed complementarity problems}
Tensor games with tensor constraints can exactly be reformulated as mixed complementarity problems by way of the Karush-Kuhn-Tucker (KKT) conditions. We use that mainstream approach here to solve them. Note that in a bimatrix game, the corresponding complementarity problem is linear, but in this is not necessarily true in $n$-player tensor games, which have multilinear terms. The full mixed complementarity problem is as follows:

\newcommand{\varmixdual}{\gamma}
\newcommand{\varunitdual}{\lambda}

\begin{equation}
    \label{eq:kkt}
    \begin{aligned}
    &\nabla_{\varprobs_i} L_i(\varprobs, \varunitdual, \varmixdual)  && \perp  0 \le \varprobs_i \le \infty, \\
    &\sum_{k=1}^{\numstrats_\idxplayer} \varprobs_{\idxplayer, k} - 1 && \perp  - \infty \le \varunitdual_\idxplayer  \leq \infty, \\
    &\varconstrainttensor_{\idxconstraint} \oprtprod \varprobs - \varconfidence_{\idxconstraint} && \perp  0 \le \varmixdual_{\idxconstraint} \le \infty, \ \forall j \in \setownership_i\\
    \end{aligned}
\end{equation}
with the Lagrangian function
\begin{equation}
    L_\idxplayer(\varprobs, \varunitdual, \varmixdual) = \varcosttensor_\idxplayer\oprtprod \varprobs - \varunitdual_i \sum_{\idxstrat = 1}^{\numstrats_\idxplayer} \varprobs_{\idxplayer, \idxstrat} - \sum_{\idxconstraint \in \setownership_\idxplayer} \varmixdual_{\idxconstraint} (\varconstrainttensor_\idxconstraint \oprtprod \varprobs - \epsilon_j)
    \label{eqn:lagrangian}
\end{equation}

\noindent for all $i \in 1..n$ and $j \in \setownership_i$. The complementarity notation $F_i(Z) \perp l_i \le Z_i \le u_i$ means either $F_i(Z) \ge 0$ and $Z_i=l_i$, or $F_i(Z) = 0$ and $l_i \le Z_i \le u_i$, or $F_i(Z) \le 0$ and $Z_i = u_i$. The second complementarity condition in Eq. \ref{eq:kkt} is simply an equality ensuring that the mixing weights sum to 1; the corresponding dual $\lambda_i$ is free. $\varmixdual_{i,j}$ are duals for the tensor constraints. 

\subsection{Using chance constraints in tensor games}
We have yet to discuss $\rho_j$, the ``constraint indicating function,'' which we used to formalize coupled constraints in Section \ref{sec:boilerplate}. Depending on the choice of $\rho_j$, constraints take on a different interpretation and different behavior. We conceptualize two possible choices in this and the next section.

\hspace{0.1cm} 

\noindent When every element of $\varconstrainttensor_{i,\idxconstraint}$ is the Boolean feasibility of the corresponding joint strategy (1 if the joint strategy is feasible, 0 if it is not), $\varconstrainttensor_{i,\idxconstraint} \oprtprod x$ is the probability with which player $i$'s constraint $\idxconstraint$ is satisfied (at any value). This is an intuitive way to constrain a mixed game, where the feasibility of the realized joint strategy is indeed probabilistic. The value $\varconfidence_{i,\idxconstraint}$ in Eq. \ref{eqn:def_tensor_game} has the interpretation of a ``confidence'', the minimum acceptable probability of an infeasible solution being realized. 

To ensure a Boolean constraint tensor when a constraint function is available, $\funsoftener(g(\varstrats_\idxtensor)) = \mathbbm{1}(g(\varstrats_\idxtensor) \geq 0)$ may be used. In practice we will find that an approximation to this is valuable.

\subsubsection{Properties}
\label{sec:chance_properties}
The primary appeal of chance constraints comes from the ability to vary $\varconfidence$ to soften constraints and remove the dependence of the solution on arbitrarily poor-quality, but feasible, pure strategies.  As such, chance constraints are applicable in ``dense'' problems with many players and significant mixing, in which many joint strategies have low feasibility due to the high number of player interactions. The existence of a solution is maintained by allowing players to mix with all strategies, not only those with no possibility of infeasibility. In this way we might expect the general amount of mixing to decrease as $\varconfidence$ increases.

On the other hand, $\varconfidence$ also has an effect on the solution quality --- low settings of it can permit low-feasibility equilibria that are filtered out at higher confidences. In games that have mixed strategy equilibria where the constraint is fully upheld, we might expect the amount of mixing to \textit{increase} with $\varconfidence$. 

At $\varconfidence_{i,\idxconstraint} = 1$, all combinations of pure strategies must be feasible w.r.t. the constraint function $g_{i,j}$, which can lead to unusual behavior. As an example, consider a game where each of two players has $\numstrats$ strategies and is incentivized to mix over all of them, with the caveat that the two players cannot choose the same strategy. At $\varconfidence_{i,\idxconstraint}=1$, as $\numstrats \rightarrow \infty$, it is impossible for all players to achieve an even mix for any finite but arbitrarily large number of lanes; there is always a (vanishingly small) likelihood of collision. 

\subsection{An Alternative: Constraints on Expectation}
\label{sec:expectation_constraints}

There is a point of view from which chance constraints appear needlessly verbose. Why not set $\rho$ to the identity and define the $K$-th element of $\varconstrainttensor_{i,j,K}$ directly as $g_{i,j}(s_K)$, in the same manner we define the elements of the cost tensor $\varcosttensor$? This is interpretable as constraining the expectation of the feasibility. They have a well-defined, intuitive meaning when the pure strategies are limited in number and static. Also, when using penalty methods (under a scalar penalty, perhaps a dynamic one), expectation-constrained games become ordinary tensor games with no irregular behavior, aside from that inherited from $\funcost$ and $\funconstraint$.  

However, expectation constraints have an unbounded dependence on the constraint function and also, (depending on that constraint function) on the strategies. As we shall see, the purpose of $\funsoftener$ is to bound this dependence. Otherwise, it is possible for changes in a pure strategy to cause arbitrarily large changes in feasibility. That is, if some pure strategy $\varstrats_{\idxplayer, \idxstrat}$ causes an extremely high feasibility, all other strategies owned by player $\idxplayer$ have no comparative effect on the feasibility. An arbitrarily small mixing weight (inversely small with respect to the feasibility value) can be assigned to $\varstrats_{\idxplayer, \idxstrat}$ and it is as though the constraint is not present.

We seek to avoid this behavior when solving continuous games, and we will return to this example to determine if this poor case in fact occurs. First, however, we develop a solver to find Nash equilibria in tensor games with tensor constraints in the following section.

        

\section{Approach}
\label{sec:implementation}

We now proceed to describe our method of solving continuous games --- choosing both the pure strategies that result in favorable cost and constraint tensors, as well as the mixing weights over these tensors. This is sometimes referred to as \textit{pure strategy optimization}, and we use a relatively ordinary approach: we append the pure strategy parameters to the decision variables of the tensor game, formulate a mixed complementarity problem using the KKT conditions, and solve. However, there are other ways of performing pure strategy optimization; for instance, \cite{peters2022lifting} formulate it as a two-level optimization problem, calculating gradients through an ordinary tensor game. We find this simple approach to be faster for the problems we explore in this work, although other approaches an interesting topic for future research.

\subsection{Solver Core}

We seek a solution to
\begin{equation}
    \forall \idxplayer \in \{1 .. \numplayers\}: \begin{bmatrix}
        \;\;\min\limits_{\varprobs_\idxplayer \in \Delta^{\numstrats_{\idxplayer - 1}}, \varstrats_\idxplayer \in \setdiscretestrats_\idxplayer^{\numstrats_\idxplayer}}\;\; \varcosttensor_{\idxplayer}(s) \oprtprod \varprobs \;\;\;\; \textrm{s.t.}\;\;\;\;\\[8pt]
        \;\;\;\;\;\;\varconstrainttensor_{i,\idxconstraint}(s) \oprtprod x \geq \varconfidence_{i,\idxconstraint} \;\; \forall \idxconstraint \in \setownership_i \\[8pt]
    \end{bmatrix}
    \label{eq:tg_aug}
\end{equation}
where $\varcosttensor_\idxplayer(s)$ and $\varconstrainttensor_{i,\idxconstraint}(s)$ are those in Eqn. \ref{eqn:def_lift_tensor_elements}.

Under differentiable $\funcost$, $\funconstraint$, and $\funsoftener$, the KKT conditions of this formulation can be taken, forming a new mixed complementarity problem where the cost and constraint tensors are functions of the strategies:

\begin{equation}
    \label{eq:kkt_aug}
    \begin{aligned}
    \nabla_{\varprobs_i} L_i(\varprobs, \varstrats, \varunitdual, \varmixdual)  & \perp 0 \le \varprobs_i \le \infty \\
    \nabla_{\varstrats_i} L_i(\varprobs, \varstrats, \varunitdual, \varmixdual)  & \perp -\infty \leq \varstrats_i \leq \infty\\
    \sum_{k=1}^{\numstrats_\idxplayer} \varprobs_{\idxplayer, k} - 1                    & \perp - \infty \leq \varunitdual_\idxplayer  \leq \infty \\
    \varconstrainttensor_{i,\idxconstraint}(\varstrats) \oprtprod \varprobs - \varconfidence_{i,\idxconstraint}      & \perp 0 \le \varmixdual_{i,\idxconstraint} \le \infty, \ \forall j \in \setownership_i, \\
    \end{aligned}
\end{equation}
with the Lagrangian function
\begin{equation}
    L_\idxplayer(\varprobs, \varstrats, \varunitdual, \varmixdual) = \varcosttensor_\idxplayer(\varstrats)\oprtprod \varprobs - \varunitdual_i \sum_{\idxstrat = 1}^{\numstrats_\idxplayer} \varprobs_{\idxplayer, \idxstrat} - \sum_{\idxconstraint \in \setownership_\idxplayer} \varmixdual_{i,\idxconstraint}(\varconstrainttensor_{i,\idxconstraint}(\varstrats) \oprtprod \varprobs).
    \label{eqn:lagrangian_aug}
\end{equation}

Unlike in the finite game setting, the optimization problems given by Eqn. \ref{eq:tg_aug} are generally non-convex with respect to the decision variables. Therefore a solution to the joint KKT conditions (\ref{eq:kkt_aug}) is not necessarily a proper equilibrium point. Nevertheless, for our purposes of exploring ways to model constraints in mixed-strategy continuous games, the KKT approach is satisfactory. The mixed complementarity problem implied by Eqn. \ref{eq:kkt_aug} is then solved with an ordinary MCP solver. The difficulty of solving this MCP depends on the nonlinearity of $f$, $g$, and $\rho$; under linear functions the problem is multilinear and therefore comparable to the MCPs defined by (\ref{eq:kkt}).

\subsection{Iterative Tightening}

The approach above only holds under twice differentiable cost, constraint, and constraint indicating functions. While it is reasonable to expect this about the cost and constraint functions in the application areas we are concerned with (robotics and control), it does not hold for $\funsoftener$ under default chance constraints, because the indicator $\mathbbm{1}$ is non-differentiable. 
To circumvent this, we ``soften'' the indicator into a sigmoid function, so that 
\begin{equation}
    \funsoftener(g(s)) = \sigma(\prmstrictness g(s)) = \frac{1}{1 + e^{-\prmstrictness g(s)}},
\end{equation}
where $\prmstrictness$ sets the extent of softening. We seek to use the highest value of $\prmstrictness$ that is numerically feasible, since  $\lim_{\prmstrictness \rightarrow \infty} \sigma(\prmstrictness x) = \mathbbm{1}(x \geq 0)$. If $\prmstrictness$ is too small, the function $\rho$ does not act as a good approximation to the indicator function, and the resulting tensor constraints will not be representative of proper chance constraints. If $\prmstrictness$ is too high, the constraint gradient will be lost numerically and the MCP will generally suffer from poor conditioning. Near the constraint boundary, this conditioning issue diminishes and a suitable solution can be found.

\begin{algorithm}
	\SetAlgoNoLine
	\KwIn{Twice differentiable cost and constraint functions $\funcost_\idxplayer$, $\funconstraint_\idxconstraint$; initial guesses $\varprobs$, $\varstrats$}
	\KwOut{Optimal strategies and mixing weights $\varstrats$ and $\varprobs$ at equilibria among the players indexed by $\idxplayer$}
    ~\\
    \For{indices $\idxtensor$ in cost tensors $\varcosttensor_\idxplayer$}{
        $\varcosttensor_{\idxplayer, \idxtensor} \leftarrow \funcost_\idxplayer(\varstrats_K)$ \;
    }
	\Repeat{$\prmstrictness \geq 2\prmstrictness_\textrm{des}$}{
         \For{indices $\idxtensor$ in constraint tensors $\varconstrainttensor_\idxconstraint$}{
             $\varconstrainttensor_{\idxconstraint, \idxtensor} \leftarrow 1 / (1 + \exp(-\prmstrictness \cdot \funconstraint_\idxconstraint(\varstrats_K)))$ \;
         }
        
         $x$, $s$ $\leftarrow$ solve\_gnep($\varcosttensor$, $\varconstrainttensor, \epsilon$; init=($\varprobs$, $\varstrats$))\;
         $\prmstrictness \leftarrow 2\prmstrictness$\;
	}
	\caption{Constrained Strategy/Weight Nash Computation}
	\label{alg:alg}
\end{algorithm}

\begin{figure*}[]
    \centering
    \includegraphics[width=0.7\textwidth, trim={0.6cm 0cm 0cm 0.8cm}]{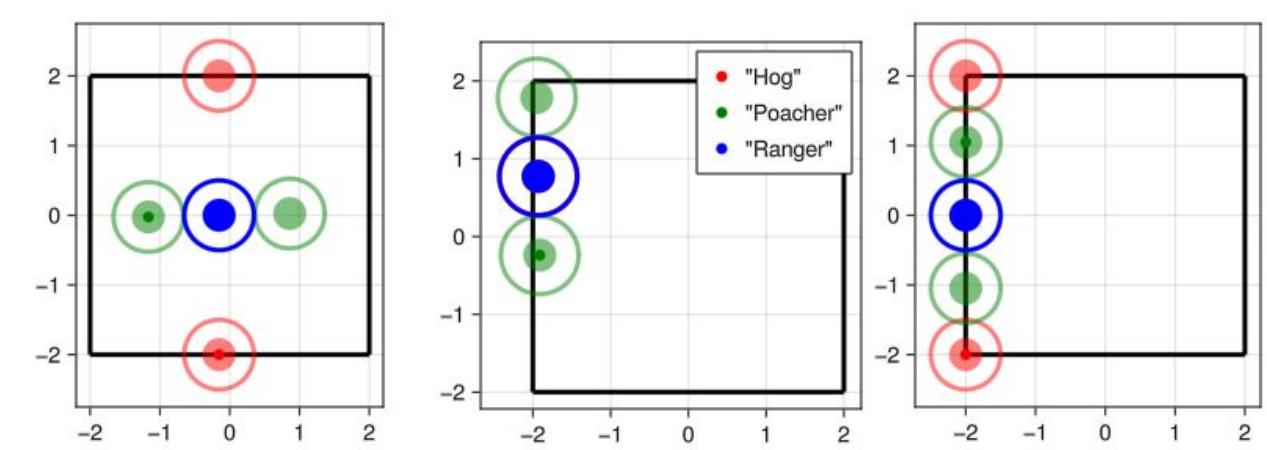}
    \caption{Potential (non-exhaustive) Nash equilibria of the Hog-Poacher-Ranger game with two pure strategies per player. Strategy weight is mapped to opacity. In (b), the ``hog'' player is rendered underneath the ``ranger'' player.}
    \label{fig:hpr_eqs}
    \Description{Three Nash equilibria of the hog-poacher-ranger game. In the first, the hog mixes 50/50 to the north and south, the ranger places itself in the middle, and the poacher mixes 50/50 to the west and east of the ranger. In the second, the hog is located under the ranger (both with full weight) and the poacher is to either side, 50/50 weight. In the third, a line is formed: 0.5 hog, 0.5 poacher, 1.0 ranger, 0.5 poacher, 0.5 hog.}
\end{figure*}

Using that insight, we apply a simple \textit{iterative tightening} procedure. We first solve the problem with a low value of  $\prmstrictness$; then, we double the strictness and solve the problem again, using the previous solution as an initialization. We repeat this process until we reach a desired strictness or the solver fails.

Algorithm \ref{alg:alg} gives the overall approach, where solve\_gnep($\varconstrainttensor, \varcosttensor, \epsilon$) formulates the KKT conditions in (\ref{eq:kkt_aug}) and solves the resulting MCP.

\subsection{Technical Details}

\subsubsection{Efficiency}
Tensor games do not scale well in the number of players as the cost tensors are exponential in size with $\numplayers$. This is a known problem, and there are a variety of methods to ameliorate it, from sampling methods \cite{gemp2021sample} to parallelization \cite{widger2009parallel} to neural network approaches \cite{marris2022turbocharging}. The addition of constraints does not further affect this complexity, as each constraint tensor is exactly the size of the cost tensors, so scaling is linear with the number of constraints.

In this paper, we assume only proper handling of sparsity in tensors, and otherwise use no efficiency modifications.

\subsubsection{Tooling}
We implement Algorithm \ref{alg:alg} in Julia \cite{bezanson2017julia}, a general-purpose mathematical programming language. We use the PATH solver \cite{dirkse1995path} to find solutions to the underlying MCPs.

\section{Analysis}

We now turn to characterizing the solver's applicability towards spatial games with intuitive collision constraints. As an illustrative scenario we consider a pursuit-evasion game. Pursuit-evader games are a useful benchmark in our considerations as in pure strategy dimension of one or higher, the evader is incentivized to mix in apparently ``deceptive'' behavior, while pure strategies themselves must be placed strategically as well \cite{peters2022lifting}. 

\subsection{Solution Fidelity and Parameter Settings}
We begin by considering the adequacy of the solver for seeking constrained equilibria in pure strategies and weights with respect to its parameters.

\newcommand{\hprdistance}{r}

\subsubsection{The ``Hog-Poacher-Ranger'' game} 

Consider a constrained pursuit-evasion game for use in our discussions of solver fidelity. In this ``hunt-in-a-hunt'' game, three players (the  ``hog'', the ``poacher'', and the ``ranger'', indexed in that order) interact in a two-dimensional space:
    \begin{itemize}
        \item The ``hog'' maximizes its squared distance from the ``poacher''.
        \item The ``poacher'' minimizes its squared distance to the ``hog'', under the constraint that it must not be within a certain distance $\hprdistance$ of the ``ranger''.
        \item The ``ranger'' minimizes its squared distance to the ``poacher''.
    \end{itemize}

    This is a tensor game with a single tensor constraint (respected only by the ``poacher'' player). Each player mixes over $\numstrats_\idxplayer$ pure strategies $\varstrats_\idxplayer$, which are also to be selected optimally. Each strategy $\varstrats_{\idxplayer, \idxstrat} \in [-2, 2]^2$. The cost and constraint functions are as follows, where $\hat{\varstrats_\idxplayer}$ is the strategy realized by player $\idxplayer$:

    \begin{equation}
    \begin{aligned}
            \funcost_1(\hat{\varstrats}) = -||\hat{\varstrats}_1 - \hat{\varstrats}_2||_2^2 \;\;\;
    \funcost_2(\hat{\varstrats}) &= ||\hat{\varstrats}_1 - \hat{\varstrats}_2||_2^2 \;\;\;\;
    \funcost_3(\hat{\varstrats}) = ||\hat{\varstrats}_2 - \hat{\varstrats}_3||_2^2 \\
    \funconstraint_1(\hat{\varstrats}) &= ||\hat{\varstrats}_2 - \hat{\varstrats}_3||_2^2 - \hprdistance^2
    \end{aligned}
    \end{equation}
    

The hog-poacher-ranger game has multiple optimal settings of the pure strategies $\varstrats$ (that is, multiple Nash equilibria in $\varprobs$ and $\varstrats$). In general, mixed strategies are used by all players except the ``ranger.'' In the remainder of this paper we allot all players of this game two pure strategies over which to mix (regardless of whether they actually must do so at equilibrium) and use $\hprdistance = 1$.

\subsubsection{Solving Hog-Poacher-Ranger}
We deployed an implementation of the solver described in Alg. \ref{alg:alg} on one thousand randomized initial solutions of the ``hog-poacher-ranger'' game for several parameter variations of chance constraints. We noted cases where the game was unable to be solved, as well as the average actual chance of violating the constraint, which generally differs from the constraint as computed during solving. We also note the cost of every player (note that the game is zero-sum between the ``hog'' and ``poacher'') and the time taken to solve. Table \ref{tab:bulk_results} summarizes these results. We also include the behavior of expectation constraints as described in Section \ref{sec:expectation_constraints}.

\begin{table*}[]
\centering
\begin{tabular}{lrrrr}
                                 & \multicolumn{1}{l}{\textbf{Chance; $\epsilon=$50\%}} & \multicolumn{1}{l}{\textbf{Chance; $\epsilon=$80\%}} & \multicolumn{1}{l}{\textbf{Chance; $\epsilon=$99\%}} & \multicolumn{1}{l}{\textbf{Expectation}} \\
\textbf{\% Solved}               & 90.70\%                                  & 83.0\%                                   & 73.1\%                                   & 98.50\%                                  \\
\textbf{Hog} (neg. Poacher) \textbf{Cost} & -3.261                                   & -3.003                                   & -4.319                                   & -3.251                                   \\
\textbf{Ranger Cost}             & 1.029                                    & 1.504                                    & 2.585                                    & 1.000                                    \\
\textbf{Feasibility Chance}  & 44.04\%                                  & 100.00\%                                 & 100.00\%                                 & 21.04\%                                  \\
\textbf{Avg Solver Time} (ms)    & 12.14                                    & 18.68                                    & 15.61                                    & 5.77                                    
\end{tabular}%
\caption{Bulk results of playing the ``hog-poacher-ranger'' game. ``Feasibility Chance'' is the likelihood that the constraint is upheld after solving, applying the exact constraint to the solution.}
\label{tab:bulk_results}
\end{table*}

\subsubsection{Parameters: Confidence}
\begin{figure}
    \centering
    \includegraphics[width=9cm,trim={1cm 1.4cm -1cm 0.8cm},clip]{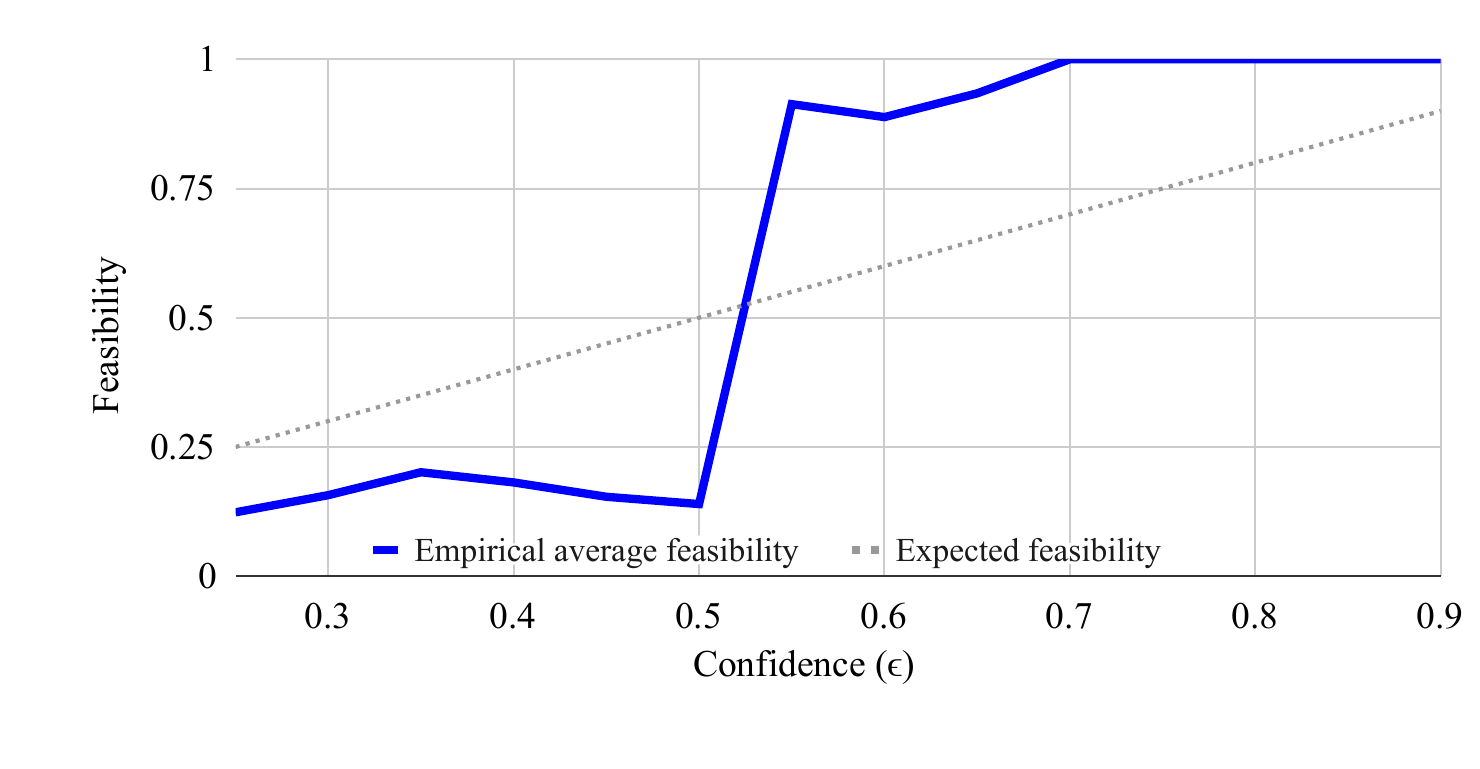}
    \caption{Feasibility at equilibrium for solves of the ordinary ``hog-poacher-ranger'' game using Alg. \ref{alg:alg}.}
    \Description{A sigmoid-shaped line plot of achieved average empirical feasibility for different settings of $\epsilon$. The theoretical feasibility is shown as the dotted line where $\epsilon$ and the realized feasibility are the same. The true, sigmoid-shaped feasibility differs from this most at $\epsilon=0.5$.}
    \label{fig:chance_eps_progression}
\end{figure}

Our application of chance constraints uses two types of parameter: $\varconfidence_{i,\idxconstraint}$ and $\prmstrictness$. We explore the impact they have on the resulting game, considering first $\varconfidence_{i,\idxconstraint}$.  Figure \ref{fig:chance_eps_progression} shows the average number of times the constraint was actually upheld in the ``hog-poacher-ranger'' game with respect to different settings of $\varconfidence_{i,\idxconstraint}$ (that is, the percentage of games in which the poacher ``escapes''). We averaged across one hundred games per setting and $\prmstrictness_\textrm{des}=64$ was used; games in which $\omega < 10$ after iterative tightening were considered solver failures. The dotted line corresponds to the expected percentage of ``escapes'' (the exact constraint). We see that settings above $\varconfidence_{i,\idxconstraint}=0.5$ the solution tends to be overly conservative, while below $\varconfidence_{i,\idxconstraint}=0.5$, the solution tends to be loose. 

We can rationalize this behavior as being caused by the limited number of mixed strategies available. In this problem, at any equilibrium, the poacher mixes their two strategies with weights [0.5, 0.5] and the ranger does not mix. Therefore, a feasibility which is not in $\{0.0, 0.5, 1.0\}$ is impossible with a perfectly strict constraint (as the poacher can mix zero, one, or two strategies). However, during solving, only the soft constraint is available, and $\varconfidence$ determines where strategies lie on it. At $\epsilon < 0.5$, pure strategies may lie slightly inside (violating) the true constraint; at $\epsilon > 0.5$ they must lie slightly outside it. When the perfectly strict version of the constraint is then used these two cases appear to have opposite feasibility, even when they may both lie very close to the constraint boundary. 

\subsubsection{Parameters: Strictness}

Based on the behavior of $\epsilon$ we expect $\omega$ to bring the pure strategies closer to the true constraint as it is increased. We investigate this further by investigating each step of the iterative tightening procedure. For each of one hundred initial positions per setting of $\epsilon$, we attempt Alg. \ref{alg:alg} on each with $\omega_\textrm{des} = 64$. We track the minimum realizable distance between the ``poacher'' an the ``ranger'' as a proxy to the fidelity with which the constraint is upheld (or overshot) for each $\omega$.

\begin{figure}
    \centering
    \includegraphics[width=8.5cm]{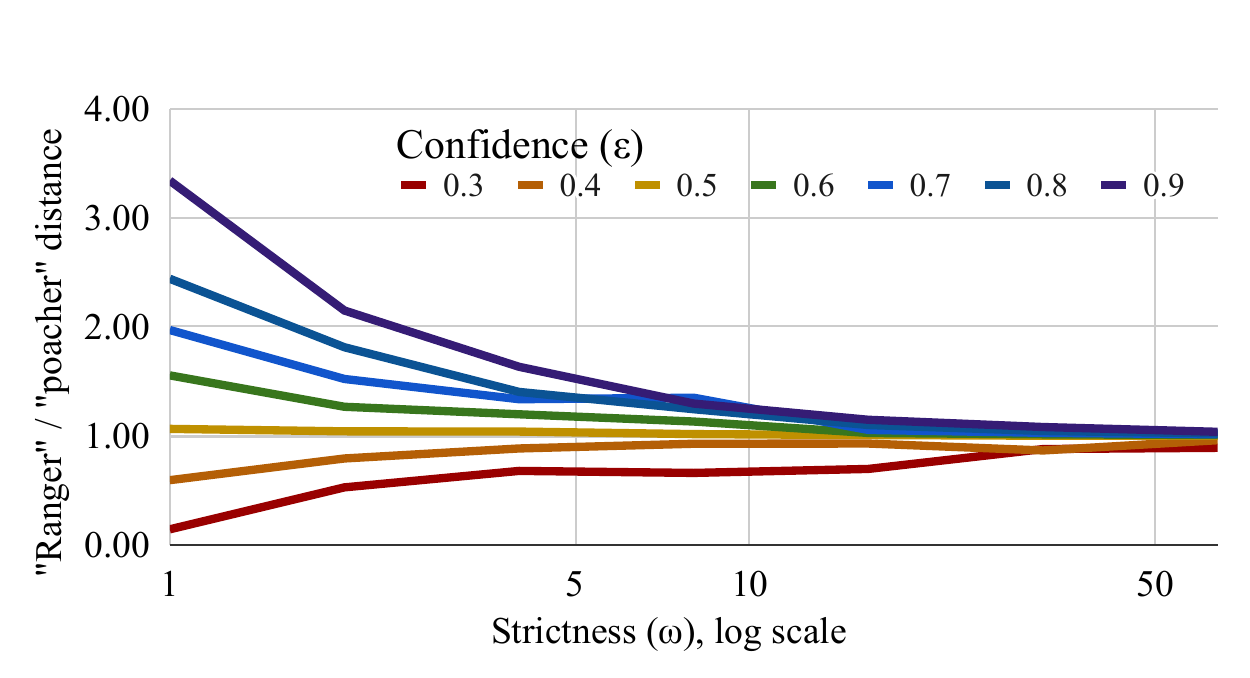}
    \caption{Effect of $\omega$ on constraint value (minimal distance between ``ranger'' and ``poacher''). Actual constraint at distance of 1.}
    \Description{Visualization of the true distance between the ranger and the poacher in the HPR game, for different $\omega$ and $\epsilon$. All converge to the correct value as $\omega$ increases. Higher $\epsilon$ values show steeper curves in this convergence.}
    \label{fig:omega_dist}
\end{figure}

Figure \ref{fig:omega_dist} shows the results. As we should expect, as $\omega$ increases, the constraint indeed becomes closer to the constraint, and the extent to which this occurs depends on the severity of $\epsilon$. When $\epsilon < 0.5$, the pure strategies converge to the constraint boundary from the inside of the constraint, which explains the unexpected relationship between confidence and true feasibility showcased in Fig. \ref{fig:chance_eps_progression}.

\subsubsection{Computation Results: N-Player Hog-Poacher-Ranger Game} To benchmark our solver, we also used an $N$-player variant of this game, in which the first player acts as the hog, the $N-2$ subsequent players act as poachers aiming to minimize expected squared distance to their predecessor, and constrained to maintain a minimum distance from their successor. The final player also aims to minimize the expected squared distance to its predecessor. Because this game is well-defined for $N$ players, we can perform a numerical analysis to evaluate the solution time when solving using Algorithm 1 for various values of $N$ as well as various values of $m$, where $m$ is the number of pure strategies held by each of the players. The results of this experiment are listed in Table \ref{tab:timing}.

\begin{table}[H]
\begin{center}
\begin{tabular}{lllll}
             & \textbf{m=1} & \textbf{m=2} & \textbf{m=3} & \textbf{m=4}  \\ \hline
\textbf{N=2} & 3.4 ± 0.5    & 8.8 ± 0.3    & 9.2 ± 0.4    & 19.2 ± 0.6    \\
\textbf{N=3} & 15.9 ± 0.7   & 53.1 ± 1.0   & 52.1 ± 1.1   & 221.3 ± 3.0   \\
\textbf{N=4} & 48.4 ± 9.2   & 69.2 ± 0.8   & 251.4 ± 2.5  & 237.6 ± 1.0   \\
\textbf{N=5} & 59.6 ± 1.1   & 476.8 ± 6.7  & 369.2 ± 4.3  & 594.0 ± 1.8   \\
\textbf{N=6} & 94.6 ± 2.2   & 418.8 ± 0.8  & 616.6 ± 2.7  & 1340.9 ± 2.4  \\
\textbf{N=7} & 84.7 ± 1.5   & 994.2 ± 1.7  & 1198.0 ± 6.0 & 2839.5 ± 14.9
\end{tabular}
\caption{Timing benchmarks (in milliseconds) on the hog-poacher-ranger game.}
\label{tab:timing}
\end{center}
\end{table}

\FloatBarrier

\subsection{Outlier Sensitivity of Expectation Constraints}
As expected, the behavior explained in Section \ref{sec:expectation_constraints} causes poor performance: selecting one extreme-feasibility strategy with low probability allows low-cost, infeasible strategies to be used with high probability. When pure strategies are continuous and subject to optimization, it is always possible to find such a strategy, exacerbating the problem in this setting. 

\begin{figure}
    \centering
    \includegraphics[width=6cm]{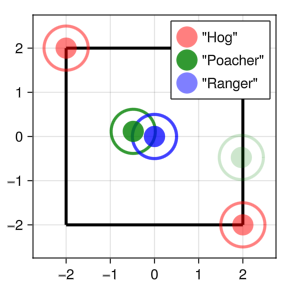}
    \caption{One solution of the ``hog-poacher-ranger'' game under expectation constraints. Strategy weight is mapped to opacity.} 
    \Description{The ``hog'' player is depicted in the far corners of the square play area, each corner at 50 percent weight. The ``ranger'' player sits between them at full weight. The ``poacher'' player has two pure strategies: one is low-weight and far away from the ranger, and the other intersects the ranger.}
    
    \label{fig:expectation_hpr_poacher_caught}
\end{figure}

For instance, in the ``hog-poacher-ranger'' example, the poacher may select pure strategies from the space $[-2, 2]^2$, and in fact often selects a pure strategy on the border that allows the ``primary'' strategy to violate the constraint. Figure \ref{fig:expectation_hpr_poacher_caught} showcases this common behavior when using expectation-based constraints.

This is not, strictly speaking, a solver failure, but rather a modeling mistake: the expected poacher-ranger distance correctly lies against the constraint (expected distance of 1). In fact, since expectation constraints are generally well-conditioned and no softening or additional parameters are required, solving this formulation is quick and rarely fails. However, expectation constraints defy our notion of ``constraints''; they provide us no guarantees on the behavior under any particular realized strategies. We would like to be able to weigh between the feasibility of every particular pure strategy, and the feasibility of the strategy set as a whole, with some parameter of the constraint.

\begin{figure}
    \centering
    \includegraphics[width=8cm]{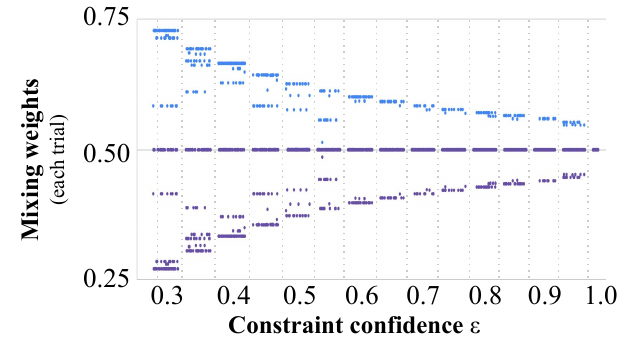}
    \caption{Mixing weights of the ``poacher'' player for each constraint confidence trial in the ``hog-poacher-ranger'' game. (The horizontal axis is discrete; points are arranged laterally for each $\varconfidence$ for visibility.)}
    \label{fig:eps_weights}
\end{figure}

In Section \ref{sec:chance_properties}, we surmised that chance constraints with a higher confidence could either prevent mixing (on the basis that most joint strategies can be infeasible and mixes are less likely to satisfy the constraint under high confidence), or promote it (on the basis that games which really do have a completely feasible mixed strategy solution are more likely to find it.) To build intuition on this matter for continuous games, we plot the actual ``poacher'' equilibrium weights achieved for each of one hundred trials across settings of $\epsilon$ in Figure \ref{fig:eps_weights}.  

\subsection{Promotion of mixed strategies}

We find that the ``hog-poacher-ranger'' game typically falls in the latter category: its equilibria are generally perfectly mixed for the ``poacher'' player, varying from this only at sufficiently low $\varconfidence_{\idxplayer,j}$. 

In general, correlating the findings of Figs. \ref{fig:eps_weights} and \ref{fig:chance_eps_progression}, we find that this game is not sensitive to settings of $\epsilon$ given a sufficient $\omega$. However, this is not necessarily the case for all games. In a version of the ``hog-poacher-ranger'' game where the poacher is allowed to mix over three strategies, at $\epsilon < 1$ equilibria exist where the poacher player overlaps the ranger player with probability $1-\epsilon$ and mixes equally over the remaining two strategies. Table \ref{tab:three_strat_eps} shows this relationship between $\epsilon$ and the realized feasibility on this variant. 

\begin{table}
\begin{tabular}{rl}
\multicolumn{1}{l}{\textbf{Confidence ($\epsilon$)}} & \textbf{Realized Feasibility} \\
0.2                                                  & 0.1992                        \\
0.3                                                  & 0.4043                        \\
0.4                                                  & 0.5901                        \\
0.5                                                  & 0.6528                        \\
0.6                                                  & 0.7013                        \\
0.7                                                  & 0.7536                        \\
0.8                                                  & 0.8062                        \\
0.9                                                  & 0.8797                       
\end{tabular}
\caption{Relationship between the confidence value $\epsilon$ and the realized (exact) feasibility at equilibrium for a modified ``hog-poacher-ranger''.}
\label{tab:three_strat_eps}
\end{table}

As we would expect, here, we do not see an immediate shift in realized feasibility at $\epsilon=0.5$ as we do in Fig. \ref{fig:chance_eps_progression}. Now, with one pure strategy not on the constraint boundary, the confidence $\epsilon$ becomes relevant to mixing. This is both a feature and a drawback of chance constraints: When many mixed strategies are present, this effect of $\epsilon$ helps prevent low-probability strategies from interfering and preventing the game from being solved, but doing so can (as in this case) obscure the true, exactly constrained equilibria, if there are any.

\FloatBarrier

\section{Closing Remarks}
\subsection{Extensions}
\subsubsection{Fairness}
In general, tensor games with tensor constraints can admit multiple solutions. This can occur the same constraint is repeated under different ownership --- for instance, $\setownership_1 =\setownership_2$, $Q_{1,j} = Q_{2,j}$. This is generally undesirable in problems in physical space (particularly with regards to collision constraints in which, loosely speaking, we expect all players to ``push'' equally against the constraints). Ideally we seek a \textit{fair} equilibria \cite{rabin1993fairness}. 

When constraints appear in the same ownership sets, the formulation used here in fact yields a \textit{refinement} of generalized Nash equilibria known as the variational equilibrium. The variational equilibrium is a sort of ``fair'' equilibrium, in that players share dual variables $\varmixdual_{\idxconstraint}$ for any particular constraint (as in \cite{schiro2013sharing} and \cite{nabetani2011sharing}). Unfortunately, this refinement is not sufficient to ensure a unique solution for tensor games with tensor constraints.

Here, we ignore this characteristic, assuming is generally the case that $\varprobs_\idxplayer > 0$. Nevertheless, it is important to remain aware of this possible redundancy to avoid unexpected complications during further optimization (i.e., over pure strategies). Furthermore, optimal pure strategy selections are generally not unique either, implying that there may be further-reaching fairness effects beyond the mixing weights alone.

\subsubsection{Information Structure}
Chance constraints can be applied to tensor games in bilevel structures. In this situation, upper-level (leader) players can intentionally spread their pure strategies and mix more evenly to block lower-level players using these constraints. Furthermore, the confidence $\varconfidence$ is compelling as a knob for controlling the extent to which the upper level player may control the lower level through the constraint. As information structure becomes complicated beyond bilevel problems, these dynamics could prove quite nuanced indeed, calling for extensive tooling to handle feasible set interactions numerically.   

\subsection{Conclusion}
We have presented a method for constraining equilibrium problems taking place in both pure strategies and mixing weights using a chance constraint based method. From the point of view of Nash equilibrium users --- specifically roboticists, control theorists, and motion planners --- the ability to handle feasible set interactions without sacrificing either strategic mixing or pure strategy optimality is critical to modeling problems in physical space. We look forward to applying this work to more complex spatial games and anticipate its utility in modeling robot interactions in game theoretic settings.




\bibliographystyle{ACM-Reference-Format} 
\bibliography{bibliography}


\begin{thebibliography}{24}


\ifx \showCODEN    \undefined \def \showCODEN     #1{\unskip}     \fi
\ifx \showDOI      \undefined \def \showDOI       #1{#1}\fi
\ifx \showISBNx    \undefined \def \showISBNx     #1{\unskip}     \fi
\ifx \showISBNxiii \undefined \def \showISBNxiii  #1{\unskip}     \fi
\ifx \showISSN     \undefined \def \showISSN      #1{\unskip}     \fi
\ifx \showLCCN     \undefined \def \showLCCN      #1{\unskip}     \fi
\ifx \shownote     \undefined \def \shownote      #1{#1}          \fi
\ifx \showarticletitle \undefined \def \showarticletitle #1{#1}   \fi
\ifx \showURL      \undefined \def \showURL       {\relax}        \fi
\providecommand\bibfield[2]{#2}
\providecommand\bibinfo[2]{#2}
\providecommand\natexlab[1]{#1}
\providecommand\showeprint[2][]{arXiv:#2}

\bibitem[\protect\citeauthoryear{Ba and Pang}{Ba and Pang}{2022}]%
        {ba2022exact}
\bibfield{author}{\bibinfo{person}{Qin Ba} {and} \bibinfo{person}{Jong-Shi Pang}.} \bibinfo{year}{2022}\natexlab{}.
\newblock \showarticletitle{Exact penalization of generalized Nash equilibrium problems}.
\newblock \bibinfo{journal}{\emph{Operations Research}} \bibinfo{volume}{70}, \bibinfo{number}{3} (\bibinfo{year}{2022}), \bibinfo{pages}{1448--1464}.
\newblock


\bibitem[\protect\citeauthoryear{Bezanson, Edelman, Karpinski, and Shah}{Bezanson et~al\mbox{.}}{2017}]%
        {bezanson2017julia}
\bibfield{author}{\bibinfo{person}{Jeff Bezanson}, \bibinfo{person}{Alan Edelman}, \bibinfo{person}{Stefan Karpinski}, {and} \bibinfo{person}{Viral~B Shah}.} \bibinfo{year}{2017}\natexlab{}.
\newblock \showarticletitle{Julia: A fresh approach to numerical computing}.
\newblock \bibinfo{journal}{\emph{SIAM review}} \bibinfo{volume}{59}, \bibinfo{number}{1} (\bibinfo{year}{2017}), \bibinfo{pages}{65--98}.
\newblock


\bibitem[\protect\citeauthoryear{Charnes and Cooper}{Charnes and Cooper}{1959}]%
        {charnes1959chancedef}
\bibfield{author}{\bibinfo{person}{Abraham Charnes} {and} \bibinfo{person}{William~W Cooper}.} \bibinfo{year}{1959}\natexlab{}.
\newblock \showarticletitle{Chance-constrained programming}.
\newblock \bibinfo{journal}{\emph{Management science}} \bibinfo{volume}{6}, \bibinfo{number}{1} (\bibinfo{year}{1959}), \bibinfo{pages}{73--79}.
\newblock


\bibitem[\protect\citeauthoryear{Dirkse and Ferris}{Dirkse and Ferris}{1995}]%
        {dirkse1995path}
\bibfield{author}{\bibinfo{person}{Steven~P Dirkse} {and} \bibinfo{person}{Michael~C Ferris}.} \bibinfo{year}{1995}\natexlab{}.
\newblock \showarticletitle{The path solver: a nommonotone stabilization scheme for mixed complementarity problems}.
\newblock \bibinfo{journal}{\emph{Optimization methods and software}} \bibinfo{volume}{5}, \bibinfo{number}{2} (\bibinfo{year}{1995}), \bibinfo{pages}{123--156}.
\newblock


\bibitem[\protect\citeauthoryear{Dutang}{Dutang}{2013}]%
        {dutang2013existence}
\bibfield{author}{\bibinfo{person}{Christophe Dutang}.} \bibinfo{year}{2013}\natexlab{}.
\newblock \showarticletitle{Existence theorems for generalized Nash equilibrium problems: An analysis of assumptions}.
\newblock \bibinfo{journal}{\emph{HAL Open Science}} (\bibinfo{year}{2013}).
\newblock


\bibitem[\protect\citeauthoryear{Facchinei and Kanzow}{Facchinei and Kanzow}{2010a}]%
        {facchinei2010generalized}
\bibfield{author}{\bibinfo{person}{Francisco Facchinei} {and} \bibinfo{person}{Christian Kanzow}.} \bibinfo{year}{2010}\natexlab{a}.
\newblock \showarticletitle{Generalized Nash equilibrium problems}.
\newblock \bibinfo{journal}{\emph{Annals of Operations Research}} \bibinfo{volume}{175}, \bibinfo{number}{1} (\bibinfo{year}{2010}), \bibinfo{pages}{177--211}.
\newblock


\bibitem[\protect\citeauthoryear{Facchinei and Kanzow}{Facchinei and Kanzow}{2010b}]%
        {facchinei2010penalty}
\bibfield{author}{\bibinfo{person}{Francisco Facchinei} {and} \bibinfo{person}{Christian Kanzow}.} \bibinfo{year}{2010}\natexlab{b}.
\newblock \showarticletitle{Penalty methods for the solution of generalized Nash equilibrium problems}.
\newblock \bibinfo{journal}{\emph{SIAM Journal on Optimization}} \bibinfo{volume}{20}, \bibinfo{number}{5} (\bibinfo{year}{2010}), \bibinfo{pages}{2228--2253}.
\newblock


\bibitem[\protect\citeauthoryear{Geletu, Kl{\"o}ppel, Zhang, and Li}{Geletu et~al\mbox{.}}{2013}]%
        {geletu2013chanceapplications}
\bibfield{author}{\bibinfo{person}{Abebe Geletu}, \bibinfo{person}{Michael Kl{\"o}ppel}, \bibinfo{person}{Hui Zhang}, {and} \bibinfo{person}{Pu Li}.} \bibinfo{year}{2013}\natexlab{}.
\newblock \showarticletitle{Advances and applications of chance-constrained approaches to systems optimisation under uncertainty}.
\newblock \bibinfo{journal}{\emph{International Journal of Systems Science}} \bibinfo{volume}{44}, \bibinfo{number}{7} (\bibinfo{year}{2013}), \bibinfo{pages}{1209--1232}.
\newblock


\bibitem[\protect\citeauthoryear{Gemp, Savani, Lanctot, Bachrach, Anthony, Everett, Tacchetti, Eccles, and Kram{\'a}r}{Gemp et~al\mbox{.}}{2021}]%
        {gemp2021sample}
\bibfield{author}{\bibinfo{person}{Ian Gemp}, \bibinfo{person}{Rahul Savani}, \bibinfo{person}{Marc Lanctot}, \bibinfo{person}{Yoram Bachrach}, \bibinfo{person}{Thomas Anthony}, \bibinfo{person}{Richard Everett}, \bibinfo{person}{Andrea Tacchetti}, \bibinfo{person}{Tom Eccles}, {and} \bibinfo{person}{J{\'a}nos Kram{\'a}r}.} \bibinfo{year}{2021}\natexlab{}.
\newblock \showarticletitle{Sample-based approximation of Nash in large many-player games via gradient descent}.
\newblock \bibinfo{journal}{\emph{arXiv preprint arXiv:2106.01285}} (\bibinfo{year}{2021}).
\newblock


\bibitem[\protect\citeauthoryear{Harker}{Harker}{1991}]%
        {harker1991generalized}
\bibfield{author}{\bibinfo{person}{Patrick~T. Harker}.} \bibinfo{year}{1991}\natexlab{}.
\newblock \showarticletitle{Generalized Nash games and quasi-variational inequalities}.
\newblock \bibinfo{journal}{\emph{European Journal of Operational Research}} \bibinfo{volume}{54}, \bibinfo{number}{1} (\bibinfo{year}{1991}), \bibinfo{pages}{81--94}.
\newblock
\showISSN{0377-2217}
\urldef\tempurl%
\url{https://doi.org/10.1016/0377-2217(91)90325-P}
\showDOI{\tempurl}


\bibitem[\protect\citeauthoryear{Henrion and Strugarek}{Henrion and Strugarek}{2008}]%
        {henrion2008chanceconvexity}
\bibfield{author}{\bibinfo{person}{Ren{\'e} Henrion} {and} \bibinfo{person}{Cyrille Strugarek}.} \bibinfo{year}{2008}\natexlab{}.
\newblock \showarticletitle{Convexity of chance constraints with independent random variables}.
\newblock \bibinfo{journal}{\emph{Computational Optimization and Applications}} \bibinfo{volume}{41}, \bibinfo{number}{2} (\bibinfo{year}{2008}), \bibinfo{pages}{263--276}.
\newblock


\bibitem[\protect\citeauthoryear{Kanzow and Steck}{Kanzow and Steck}{2016}]%
        {kanzow2016augmented}
\bibfield{author}{\bibinfo{person}{Christian Kanzow} {and} \bibinfo{person}{Daniel Steck}.} \bibinfo{year}{2016}\natexlab{}.
\newblock \showarticletitle{Augmented Lagrangian methods for the solution of generalized Nash equilibrium problems}.
\newblock \bibinfo{journal}{\emph{SIAM Journal on Optimization}} \bibinfo{volume}{26}, \bibinfo{number}{4} (\bibinfo{year}{2016}), \bibinfo{pages}{2034--2058}.
\newblock


\bibitem[\protect\citeauthoryear{Krawczyk}{Krawczyk}{2007}]%
        {krawczyk2007numerical}
\bibfield{author}{\bibinfo{person}{Jacek Krawczyk}.} \bibinfo{year}{2007}\natexlab{}.
\newblock \showarticletitle{Numerical solutions to coupled-constraint (or generalised Nash) equilibrium problems}.
\newblock \bibinfo{journal}{\emph{Computational Management Science}}  \bibinfo{volume}{4} (\bibinfo{year}{2007}), \bibinfo{pages}{183--204}.
\newblock


\bibitem[\protect\citeauthoryear{Marris, Gemp, Anthony, Tacchetti, Liu, and Tuyls}{Marris et~al\mbox{.}}{2022}]%
        {marris2022turbocharging}
\bibfield{author}{\bibinfo{person}{Luke Marris}, \bibinfo{person}{Ian Gemp}, \bibinfo{person}{Thomas Anthony}, \bibinfo{person}{Andrea Tacchetti}, \bibinfo{person}{Siqi Liu}, {and} \bibinfo{person}{Karl Tuyls}.} \bibinfo{year}{2022}\natexlab{}.
\newblock \showarticletitle{Turbocharging solution concepts: Solving NEs, CEs and CCEs with neural equilibrium solvers}.
\newblock \bibinfo{journal}{\emph{Advances in Neural Information Processing Systems}}  \bibinfo{volume}{35} (\bibinfo{year}{2022}), \bibinfo{pages}{5586--5600}.
\newblock


\bibitem[\protect\citeauthoryear{Mukherjee}{Mukherjee}{1980}]%
        {mukherjee1980mixed}
\bibfield{author}{\bibinfo{person}{SP Mukherjee}.} \bibinfo{year}{1980}\natexlab{}.
\newblock \showarticletitle{Mixed strategies in chance-constrained programming}.
\newblock \bibinfo{journal}{\emph{Journal of the Operational Research Society}} \bibinfo{volume}{31}, \bibinfo{number}{11} (\bibinfo{year}{1980}), \bibinfo{pages}{1045--1047}.
\newblock


\bibitem[\protect\citeauthoryear{Nabetani, Tseng, and Fukushima}{Nabetani et~al\mbox{.}}{2011}]%
        {nabetani2011sharing}
\bibfield{author}{\bibinfo{person}{Koichi Nabetani}, \bibinfo{person}{Paul Tseng}, {and} \bibinfo{person}{Masao Fukushima}.} \bibinfo{year}{2011}\natexlab{}.
\newblock \showarticletitle{Parametrized variational inequality approaches to generalized Nash equilibrium problems with shared constraints}.
\newblock \bibinfo{journal}{\emph{Computational Optimization and Applications}} \bibinfo{volume}{48}, \bibinfo{number}{3} (\bibinfo{year}{2011}), \bibinfo{pages}{423--452}.
\newblock


\bibitem[\protect\citeauthoryear{Peters, Fridovich-Keil, Ferranti, Stachniss, Alonso-Mora, and Laine}{Peters et~al\mbox{.}}{2022}]%
        {peters2022lifting}
\bibfield{author}{\bibinfo{person}{Lasse Peters}, \bibinfo{person}{David Fridovich-Keil}, \bibinfo{person}{Laura Ferranti}, \bibinfo{person}{Cyrill Stachniss}, \bibinfo{person}{Javier Alonso-Mora}, {and} \bibinfo{person}{Forrest Laine}.} \bibinfo{year}{2022}\natexlab{}.
\newblock \showarticletitle{Learning mixed strategies in trajectory games}.
\newblock \bibinfo{journal}{\emph{arXiv preprint arXiv:2205.00291}} (\bibinfo{year}{2022}).
\newblock


\bibitem[\protect\citeauthoryear{Rabin}{Rabin}{1993}]%
        {rabin1993fairness}
\bibfield{author}{\bibinfo{person}{Matthew Rabin}.} \bibinfo{year}{1993}\natexlab{}.
\newblock \showarticletitle{Incorporating fairness into game theory and economics}.
\newblock \bibinfo{journal}{\emph{The American economic review}} (\bibinfo{year}{1993}), \bibinfo{pages}{1281--1302}.
\newblock


\bibitem[\protect\citeauthoryear{Schiro, Pang, and Shanbhag}{Schiro et~al\mbox{.}}{2013}]%
        {schiro2013sharing}
\bibfield{author}{\bibinfo{person}{Dane~A Schiro}, \bibinfo{person}{Jong-Shi Pang}, {and} \bibinfo{person}{Uday~V Shanbhag}.} \bibinfo{year}{2013}\natexlab{}.
\newblock \showarticletitle{On the solution of affine generalized Nash equilibrium problems with shared constraints by Lemke’s method}.
\newblock \bibinfo{journal}{\emph{Mathematical Programming}} \bibinfo{volume}{142}, \bibinfo{number}{1-2} (\bibinfo{year}{2013}), \bibinfo{pages}{1--46}.
\newblock


\bibitem[\protect\citeauthoryear{Singh and Lisser}{Singh and Lisser}{2018}]%
        {singh2018variational}
\bibfield{author}{\bibinfo{person}{Vikas~Vikram Singh} {and} \bibinfo{person}{Abdel Lisser}.} \bibinfo{year}{2018}\natexlab{}.
\newblock \showarticletitle{Variational inequality formulation for the games with random payoffs}.
\newblock \bibinfo{journal}{\emph{Journal of Global Optimization}}  \bibinfo{volume}{72} (\bibinfo{year}{2018}), \bibinfo{pages}{743--760}.
\newblock


\bibitem[\protect\citeauthoryear{Stein, Ozdaglar, and Parrilo}{Stein et~al\mbox{.}}{2008}]%
        {stein2008separable}
\bibfield{author}{\bibinfo{person}{Noah~D Stein}, \bibinfo{person}{Asuman Ozdaglar}, {and} \bibinfo{person}{Pablo~A Parrilo}.} \bibinfo{year}{2008}\natexlab{}.
\newblock \showarticletitle{Separable and low-rank continuous games}.
\newblock \bibinfo{journal}{\emph{International Journal of Game Theory}} \bibinfo{volume}{37}, \bibinfo{number}{4} (\bibinfo{year}{2008}), \bibinfo{pages}{475--504}.
\newblock


\bibitem[\protect\citeauthoryear{von Heusinger}{von Heusinger}{2009}]%
        {von2009numerical}
\bibfield{author}{\bibinfo{person}{Anna von Heusinger}.} \bibinfo{year}{2009}\natexlab{}.
\newblock \emph{\bibinfo{title}{Numerical methods for the solution of the generalized Nash equilibrium problem}}.
\newblock \bibinfo{thesistype}{Ph.D. Dissertation}. \bibinfo{school}{Universit{\"a}t W{\"u}rzburg}.
\newblock


\bibitem[\protect\citeauthoryear{Widger and Grosu}{Widger and Grosu}{2009}]%
        {widger2009parallel}
\bibfield{author}{\bibinfo{person}{Jonathan Widger} {and} \bibinfo{person}{Daniel Grosu}.} \bibinfo{year}{2009}\natexlab{}.
\newblock \showarticletitle{Parallel computation of nash equilibria in n-player games}. In \bibinfo{booktitle}{\emph{2009 International Conference on Computational Science and Engineering}}, Vol.~\bibinfo{volume}{1}. IEEE, \bibinfo{pages}{209--215}.
\newblock


\bibitem[\protect\citeauthoryear{Zhang, Hadji, Lisser, and Mezali}{Zhang et~al\mbox{.}}{2022}]%
        {zhang2022variational}
\bibfield{author}{\bibinfo{person}{Shangyuan Zhang}, \bibinfo{person}{Makhlouf Hadji}, \bibinfo{person}{Abdel Lisser}, {and} \bibinfo{person}{Yacine Mezali}.} \bibinfo{year}{2022}\natexlab{}.
\newblock \showarticletitle{Variational Inequality for n-Player Strategic Chance-Constrained Games}.
\newblock \bibinfo{journal}{\emph{SN Computer Science}} \bibinfo{volume}{4}, \bibinfo{number}{1} (\bibinfo{year}{2022}), \bibinfo{pages}{82}.
\newblock


\end{thebibliography}


\end{document}